# Dynamic Scaling Function at the Quasiperiodic Transition to Chaos

Ronnie Mainieri[1] and Robert E. Ecke[1,2]
Center for Nonlinear Studies[1] and Physics Division[2]
Los Alamos National Laboratory, Los Alamos, NM 87545

July 25, 1994

**Abstract**

We obtain a five-step approximation to the quasiperiodic dynamic scaling function for experimental Rayleigh-Bénard convection data. When errors are taken into account in the experiment, the $f(\alpha)$ spectrum of scalings is equivalent to just two of these five scales. To overcome this limitation, we develop a robust technique for extracting the scaling function from experimental data by reconstructing the dynamics of the experiment.

## 1   Introduction

The most complete invariant description of a chaotic dynamical system is the dynamical scaling function [1] which provides scale factors that allow the reconstruction of the dynamics and the direct evaluation of its ergodic measures. Given the scaling function one can calculate all other invariant quantities (long time averages) describing the dynamical system, such as the spectrum of singularities $f(\alpha)$ or the correlation functions. Thus, the scaling function is the quantity of choice for characterizing a particular chaotic dynamical system. Theoretically the scaling function has been computed for the period-doubling and quasiperiodic transitions to chaos. Because of the completeness of the scaling function, a comparison between theoretical and experimental scaling function is a much more rigorous test of universality than is commonly demonstrated with the $f(\alpha)$ spectrum of singularities.

Despite the importance of constructing the scaling function from experimental data, few attempts have been made and these have not been convincing [2]. The difficulty lies in the sensitivity of the scaling function to variations in parameters such as can come from experimental noise or drift and in the proper definition of the scaling function, none of which have been addressed in experimental data analysis. We explain here how to overcome these difficulties and give the first definitive comparison of experimental and theoretical scaling functions for the quasiperiodic transition to chaos.

Why have previous attempts to extract the scaling function failed? There are basically two reasons: The first is that the *universal* scaling function has to be computed not from any map, but from the universal function that satisfies the Cvitanović-Feigenbaum functional equation [3]. Experimentally the universal function is not available, but it can be approximated from its definition by considering not the map obtained from the data, but one of its iterates. The second reason is that the scaling function is very sensitive to variations in the control parameters and these variations can never be completely eliminated. Numerical studies with the sine circle map show that large variations can be expected in the scaling function if the parameters are not exactly tuned to the quasiperiodic state. Here we will show how to approximate the behavior of the universal function by considering only orbit points close to the critical point, and how variation of parameters can be controlled by reconstructing the dynamics of the system.

The circle map scaling function is a generalization of Shenker's contraction rate $\alpha$ [4] to all points in the neighborhood of the inflection point of the circle map. Shenker's $\alpha$ measures the rate of exponential contraction of the close return distances of the inflection point of the circle map. We focus on the scaling function for circle maps, as it is a common case in physical systems, arising generically when two oscillators are non-linearly coupled. We use the sine circle map for numerical comparisons and obtain experimental data from a hydrodynamical experiment: Rayleigh-Bénard convection in a $^3$He-superfluid-$^4$He mixture. This system, which closely approximates classical thermal convection, has been extensively studied in the quasiperiodic regime [5, 6, 7].

The universality theory for circle maps is of wide interest because it occurs whenever two oscillators are nonlinearly coupled (the frequency of oscillation depends on the amplitude). If the coupling is strong, the system will go chaotic, but for any coupling there will be mode-locking. This was first reported by Christian Huyghens in 1665 when he described how clocks set on a shelf would synchronize the motion of their pendula [8]. The phase space of the two oscillators is composed of their amplitudes and phases and is thus four dimensional. In general, the oscillators are dissipative and therefore the study of their long term behavior may be reduced to the study of limit cycle sets. In the simplest case one of the oscillators has a fixed frequency and is driving the other oscillator, and the phase space may be reduced to the motion on the two-dimensional surface of a torus. On this torus there is a vector field that determines the motion. If it where a harmonic oscillator, then the vector field would be a set of vectors all pointing in the same direction, as the vector field (a) in figure 1. The typical frequency of oscillation is an irrational number (the rationals have measure zero), and the system winds around the torus ergodically. As the orbit winds around, it comes arbitrarily close to any point it has already visited. If we add a perturbation to this oscillator (vector field (b) in the figure), as we do when we couple it nonlinearly to another oscillator, then whenever there is a close return of the orbit to previously visited points there is a chance that the orbit will close on itself and be a periodic orbit — the system would be mode-locked. Peixoto [9] proved that this is the general case by showing that a typical vector field on a torus is mode-locked.

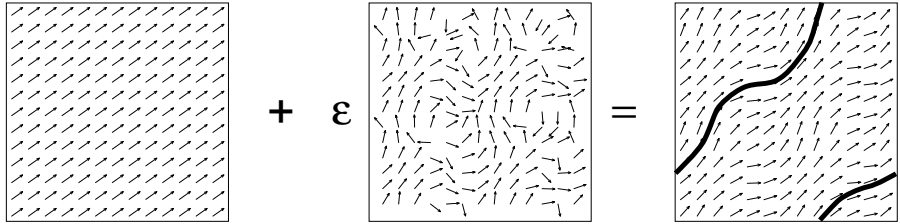

Figure 1: Mode locking occurs for typical vector fields on a torus. If a harmonic oscillator with irrational frequency is slightly perturbed, the winding trajectory will form periodic orbits.

Here we show a non-trivial physical realization of the mode-locking phenomenon. Peixoto's theorem applies to a two-dimensional dynamical system, but our analysis of the experimental data will show that the phenomenon is also observed in a system that is described by a partial differential equation (Navier-Stokes equation) and therefore potentially an infinite dimensional dynamical system. The experiment is *not* used as an analog computer to simulate equations that are known to have mode-locking. A detailed analysis of the parameter space of the experimental system shows that it is not globally equivalent to the sine circle map [5]. There are non-chaotic regions in the experiment that do not occur in a simple circle map. This is not an unusual situation for higher dimensional maps that have invariant circles (see, for example Wang *et al.* [10]).

This paper is organized by first presenting a review (section 2) of the major features of the dynamical scaling function, emphasizing connections to physically obtainable data sets (either by experiment or numerical simulation). Next, in section 3, a brief description of the experiment is given. In section 4, we present the results of our data analysis, illustrating potential difficulties using numerical simulation, but concentrating on obtaining a reliable scaling function for experimental data. Our summary and conclusions are contained in section 5.

## 2  Scaling functions

Fractals are complicated sets to describe. As a consequence several possible descriptions have been proposed, with varying degrees of completeness. The coarsest description of a fractal is its fractal dimension. It gives an idea of how many small boxes of fixed size are needed to cover the set. The closer the fractal dimension is to the embedding dimension, the closer it appears to be a figure of non-zero measure. A more detailed description of the fractal is given by the spectrum of singularities, the $f(\alpha)$ curve, or equivalently, the generalized dimensions $D_q$. The $f(\alpha)$ curve generalizes the fractal dimension by decomposing the fractal set into self-affine fractals (which are not multifractals) indexed by $\alpha$ and for each $\alpha$ gives its fractal dimension. Most fractals encountered in physics have this multitude of scales and a parabola-shaped $f(\alpha)$ curve. The $f(\alpha)$ curve has more information than the fractal dimension, as it describes the decomposition of the fractal set. Despite the infinite number of fractal dimensions it contains, it is still not possible to reconstruct the fractal from the $f(\alpha)$ curve — $f(\alpha)$ is not a complete description of the fractal. One might argue that a complete description of the fractal is not desirable, because it would be too complicated and because in principle the dynamical system that generated the fractal already provides a complete description. Such a complicated description would not be useful for comparing experiments to theories. Such an argument would be correct if there where no structure to a fractal, but there is. Fractals that occur in physical system are seldom arbitrary, and are usually described by a smooth presentation function, or equivalently a scaling function. It was Feigenbaum who observed that there is a hierarchical structure to the descriptions of a fractal that can be explored to create a function — the scaling function — which can be easily approximated by a simple function. By a simple function we mean a function that has a good approximation in terms of a basis of computable functions. For example, most $f(\alpha)$ curves can be very well approximated (to less than 1%) by a parabola, and therefore are well approximated by three numbers and the basis functions 1, $x$, and $x^2$. The scaling function would be of little practical value if it were not well approximated in a simple basis, step functions in this case.

There are many routes that lead to an explanation of what a scaling function is and how to compute it. The shortest is by breaking away from the historical development and considering first the presentation function of a fractal. The presentation function is a simple chaotic dynamical system (hyperbolic, unlike the circle map) that generates the fractal and is closely related to the definition of fractals of Hutchinson [11] and the iterated dynamical systems introduced by Barnsley and collaborators [12]. From the presentation function one can derive the scaling function, but we will not do it in the most elegant fashion, rather we will develop the formalism in a form that is directly applicable to the experimental data.

In the upper part of figure 2 we have the successive steps of the construction similar to the middle third Cantor set. The construction is done in levels, each level being formed by a collection of segments. From one level to the next, each "parent" segment produces smaller "children" segments by removing the middle section. As the construction proceeds, the segments better approximate the Cantor set. In the figure not all the segments are the same size, some are larger and some are smaller, as is the case with multifractals. In the middle third Cantor set, the ratio between a segment and the one it was generated from is exactly 1/3, but in the case shown in the figure the ratios differ from 1/3. If we went through the last level of the construction and made a plot of the segment number and its ratio to its parent segment we would have a scaling function, as indicated in the figure. A function giving the ratios in the construction of a fractal is the basic idea for a scaling function. Much of the formalism that we will introduce is to be able to give precise names to every segments and to arrange the "lineage" of segments so that the children segments have the correct parent. If we do not take these precautions, the scaling function would be a "wild function", varying rapidly and not approximated easily by simple functions.

To describe the formalism we will use a variation on the quadratic map that appears in the theory of period doubling. This is because the combi-natorial manipulations are much simpler for this map than they are for the circle map. The scaling function will be described for a one dimensional map $F$ as shown in figure 3. Drawn is the map

$$F(x) = 5x(1-x) \tag{1}$$

restricted to the unit interval. We will see that this map is also a presentation function.

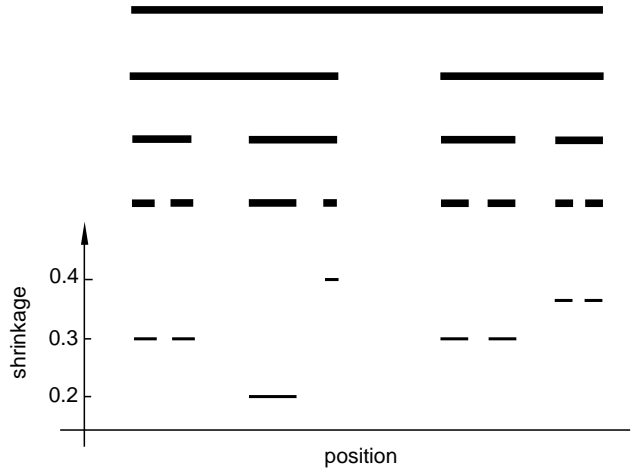

Figure 2: *Construction of the steps of the scaling function from a Cantor set. From one level to the next in the construction of the Cantor set the covers are shrunk, each parent segment into two children segments. The shrinkage of the last level of the construction is plotted and by removing the gaps one has an approximation to the scaling function of the Cantor set.*

It has two branches separated by a gap: one over the left portion of the unit interval and one over the right. If we choose a point $x$ at random in the unit interval and iterate it under the action of the map $F$, equation (1), it will hop between the branches and eventually get mapped to minus infinity. An orbit point is guaranteed to go to minus infinity if it lands in the gap. The hopping of the point defines the orbit of the initial point $x$: $x \mapsto x_1 \mapsto x_2 \mapsto \cdots$. For each orbit of the map $F$ we can associate a symbolic code. The code for this map is formed from 0s and 1s and is found from the orbit by associating a 0 if $x_t < 1/2$ and a 1 if $x_t > 1/2$, with $t = 0, 1, 2, \ldots$.

Most initial points will end up in the gap region between the two branches. We then say that the orbit point has escaped the unit interval. The points that do not escape form a Cantor set $\mathcal{C}$ (or Cantor dust) and remain trapped in the unit interval for all iterations. In the process of describing all the points that do not escape, the map $F$ can be used as a presentation of the Cantor set $\mathcal{C}$, and has been called a presentation function by Feigenbaum [13].

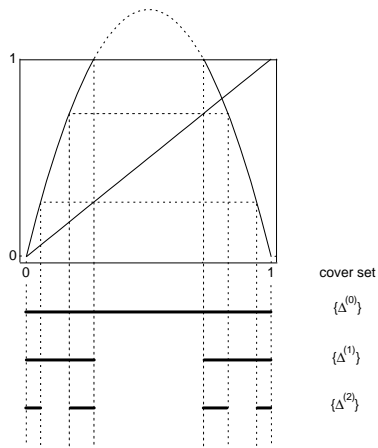

*Figure 3: A Cantor set presentation function. The Cantor set is the set of all points that under iteration do not leave the interval* [0, 1]. *This set can be found by backwards iterating the gap between the two branches of the map. The dotted lines can be used to find these backward images. At each step of the construction one is left with a set of segments that form a cover of the Cantor set.*

How does the map $F$ "present" the Cantor set? The presentation is done in steps. First we determine the points that do not escape the unit interval in one iteration of the map. These are the points that are not part of the gap. These points determine two segments, which are an approximation to the Cantor set. In the next step we determine the points that do not escape in two iterations. These are the points that get mapped into the gap in one iteration, as in the next iteration they will escape; these points form the two segments $\Delta_0^{(1)}$ and $\Delta_1^{(1)}$ at level 1 in figure 3. The processes can be continued for any number of iterations. If we observe carefully what is being done, we discover that at each step the pre-images of the gap (backward iterates) are being removed from the unit interval. As the map has two branches, every point in the gap has two pre-images, and therefore the whole gap has two pre-images in the form of two smaller gaps. To generate all the gaps in the Cantor set one just has to iterate the gap backwards. Each iteration of the gap defines a set of segments, with the $n$-th iterate defining the segments $\Delta_k^{(n)}$ at level $n$. For this map there will be $2^n$ segments at level $n$, with the first few drawn in figure 3. As $n \to \infty$ the segments that remain for at least $n$ iterates converge to the Cantor set $\mathcal{C}$.

The segments at one level form a cover for the Cantor set and it is from a cover that all the invariant information about the set is extracted (the cover generated from the backward iterates of the gap form a Markov partition for the map as a dynamical system). The segments $\{\Delta_k^{(n)}\}$ at level $n$ are a refinement of the cover formed by segments at level $n-1$. From successive covers we can compute the trajectory scaling function, the spectrum of scalings $f(\alpha)$, and the generalized dimensions.

To define the scaling function we must give labels (names) to the segments. The labels are chosen so that the definition of the scaling function allows for simple approximations. As each segment is generated from an inverse image of the unit interval, we will consider the inverse of the presentation function $F$. Because $F$ does not have a unique inverse, we have to consider restrictions of $F$. Its restriction to the first half of the segment, from 0 to 1/2, has a unique inverse, which we will call $F_0^{-1}$, and its restriction to the second half, from 1/2 to 1, also has a unique inverse, which we will call $F_1^{-1}$. For example, the segment labeled $\Delta^{(2)}(0,1)$ in figure 3 is formed from the inverse image of the unit interval by mapping $\Delta^{(0)}$, the unit interval, with $F_1^{-1}$ and then $F_0^{-1}$, so that the segment

$$\Delta^{(2)}(0,1) = F_0^{-1}\left(F_1^{-1}\left(\Delta^{(0)}\right)\right) \quad . \tag{2}$$

The mapping of the unit interval into a smaller interval is what determines its label. The sequence of the labels of the inverse maps is the label of the segment:

$$\Delta^{(n)}(\epsilon_1, \epsilon_2, \ldots, \epsilon_n) = F_{\epsilon_1}^{-1} \circ F_{\epsilon_2}^{-1} \circ \cdots F_{\epsilon_n}^{-1}\left(\Delta^{(0)}\right) \quad .$$

The scaling function is formed from a set of ratios of segments length. We use $|\cdot|$ around a segment $\Delta^{(n)}(\epsilon)$ to denote its size (length), and define

$$\sigma^{(n)}(\epsilon_1, \epsilon_2, \ldots, \epsilon_n) = \frac{|\Delta^{(n)}(\epsilon_1, \epsilon_2, \ldots, \epsilon_n)|}{|\Delta^{(n-1)}(\epsilon_2, \ldots, \epsilon_n)|} \quad .$$

We can then arrange the ratios $\sigma^{(n)}(\epsilon_1, \epsilon_2, \ldots, \epsilon_n)$ next to each other as piecewise constant segments in increasing order of their binary label $\epsilon_1, \epsilon_2, \ldots, \epsilon_n$ so that the collection of steps scan the unit interval. As $n \to \infty$ this collection of steps will converge to the scaling function. In section 4 we will describe the limiting process in more detail, and give a precise definition on how to arrange the ratios.

The construction we gave for the scaling function cannot be used for the circle map or the quadratic map (the map $ax(1-x)$, $a < 4$) because neither is hyperbolic. The essential point of the construction of Feigenbaum and collaborators was to realize that there is a way of re-writing these maps so that they are effectively hyperbolic. In both cases a universal function is constructed, and from it the scaling function or a presentation function can be computed. The universal function can be computed from the circle map $f$. Assuming that the map $f$ has an inflection point at 0, Shenker [14] observed that if we compose $f$ with itself a $Q_n$ Fibonacci number of times and choose a suitable value for $\alpha$ we can approach the universal function $g$

$$\alpha^n f^{Q_n}(\alpha^{-n} x) \to g(x) \tag{3}$$

as $x \to 0$ and $n \to \infty$. From this relation we discover that the universal function satisfies a functional equation

$$Tg(x) = \alpha g(\alpha g(\alpha^{-2} x)) \tag{4}$$

(the usual functional equation uses the function $\alpha g$). We interpret the $x \to 0$ condition as stating that the universality results hold at the origin. The $n \to \infty$ condition we interpret from the combinatorics of the circle map as meaning that we must iterate the map until the orbit lands close to the origin. For the computation of the scaling function we do not need, in principle, to restrict ourselves to the inflection point, as the scaling function is invariant under smooth diffeomorphism and also the action of the circle map itself. So if we use an image of the inflection point under the action of the map, we should be able to compute the scaling function. This would be correct if we could also take the $n \to \infty$ limit, but in practice data sets are limited. It is then no longer true that the scaling function can be computed at any image of the inflection point. A scaling function computed at an image converges more slowly than the scaling function computed at the origin. In section 4 we will give a procedure that effectively computes the scaling function from the universal function by using only iterates of the circle map.

# 3   Experiment

The experiment that provided the data for the analysis presented here has been well studied in the quasiperiodic regime using an apparatus that is described in detail elsewhere [5, 6, 7]. Here we give the essential features of the experimental data and discuss how the data are prepared for the calculation of the dynamical scaling function. The system is thermal convection in a $^3$He-superfluid-$^4$He mixture which approximates a classical convecting fluid with low Prandtl number. In some region of parameter space quasiperiodic, mode-locked, and chaotic states are observed. These states are the result of two internal oscillatory modes and not the result of external forcing. To study a quasiperiodic/mode-locking system one must vary two parameters independently. For this system, the two parameters are the temperature difference across the fluid layer and the mean temperature of the fluid. The range of rotation numbers that can be accessed by varying these parameters is from about $1/8$ to $1/6$. The canonical golden-mean rotation number, $\rho_g = (\sqrt{5} - 1)/2 \approx 13/21$ does not fall within this range but there are many rotation numbers with the proper golden mean tail (asymptotic series of 1s in rational approximant series) that are in that range. For the purposes of testing universality of the quasiperiodic transition to chaos any of these rotation numbers is equivalent. The experimental data we use is centered around the golden-mean-tail irrational $(3 - \sqrt{5})/2$. The strict golden mean has the advantage of making the "best" use of the data, as the data requirements in terms of stability and precision do increase for rotation numbers other than the strict golden mean.

The data are time sequences of temperature oscillations of the convective flow field measured at a local point in space. To the extent that this is a low dimensional dynamical system, measurement at a single point completely characterizes the state of the system. The data are used to reconstruct the phase-space dynamics using standard delay-coordinate embedding techniques [15, 16]. Poincaré sections are produced by interpolating the intersections of the phase-space trajectories with a plane. For a quasiperiodic attractor, the section will fill up a smooth curve diffemorphic with a circle. In figure 4 we show a Poincaré section for a state very near criticality (as defined in circle map descriptions of the quasiperiodic transition). Each point in the section has a time ordering according to its relative position in the time sequence and a space ordering that relates the nearest neighbor points along the Poincaré section. The correspondence between the time and space ordering is determined by the rotation number and also sets the sequence of close return times for the dynamics. Another effective action of the dynamics on this curve is to partition the curve into segments that connect nearest-neighbor points. It is these segments that form the set which obeys multifractal scaling and from which we will construct the scaling function.

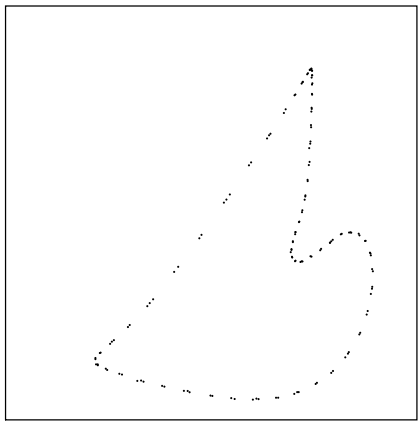

Figure 4: Poincaré cross section from the reconstructed flow.

Rather than considering the Poincaré section, we can make a further simplification by constructing a one-dimensional mapping of arc length along the curve. Such a mapping, for the data in figure 4, is shown in figure 5, and is clearly a one-dimensional map very similar to a sine circle map. For more subcritical parameters, such maps constructed from data show all the simple features of a sine circle map [7]. Practical considerations that arise for experimental data used to construct a multifractal description of the attractor (fractal dimension, $f(\alpha)$ spectrum, etc.) are the precision with which one can define a rotation number, the degree of random noise in the signal, and the stability of the state against drift in the parameters. In these experiments the signal to noise ratio was about 1000 : 1 and the rotation number could be determined to about 5 parts in a million. The most important factor which limited the data was drift in the operating parameter of the system. This caused changes in the rotation rate with time and, although very small, had extremely deleterious effects on the extraction of a scaling function for reasons discussed in the next section. In general the analysis for determining the scaling function is more demanding on the quality of the data than for averaged multifractal quantities because one is comparing the ratios of individual segments as opposed to the averaging over many segments that defines the $f(\alpha)$ spectrum. In the next section we extract a scaling function for the quasiperiodic transition to chaos by reconciling the limitations of the experimental data with the correct definition of the scaling function. In particular, the proper limits must be observed. We demonstrate these methods on numerical sine circle map data to illustrate the pitfalls of scaling function analysis.

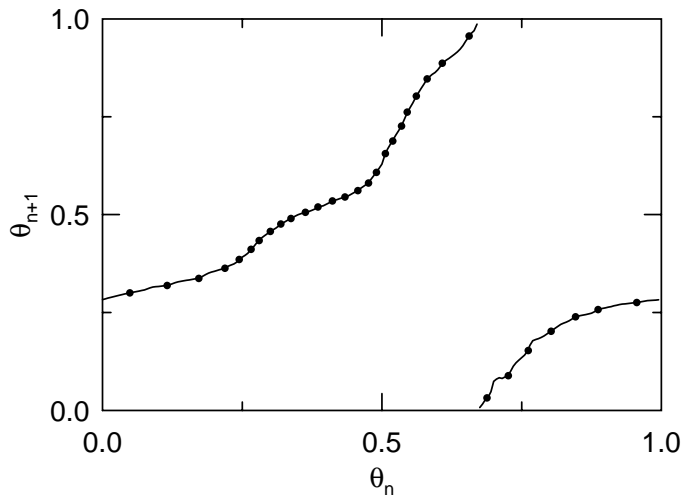

*Figure 5: Return map reconstructed from figure 4 using arc length. The dots are the data points and the solid curve the reconstructed dynamics as explained at the end of section 4.*

## 4 The practice of circle map scaling functions

In this section we will use the concepts developed earlier about scaling functions and adapt them to the requirements of circle maps and the realities of experimental data. We will have two types of difficulties. The simplest to overcome are those related to the combinatorics of the circle map, as it is not as simple as that of period-doubling maps. In the case of the circle map at a golden-mean-tail rotation number, not every segment gets sub-divided into two children segments; some do and others do not. The definition we adopt here for the scaling function matches that used by Feigenbaum in his presentation function article for circle maps [3], except that we use forward iterates. The other type of difficulties are associated with trying to deter-mine the exact point where the data should be collected. We will conclude that it is not possible to collect the data reliably at the irrational rotation number and will therefore reconstruct the dynamics of the system as a function of one parameter in the vicinity of the irrational rotation number.

In a numerical or laboratory experiment, owing to finite precision, the winding number is never an irrational, and the best that can be obtained is a rational approximant. In this case the map is locked into a periodic orbit and the range of parameters that have this frequency form a section of an Arnold tongue. The approximants are formed from truncations of the continued fraction expansion of the irrational winding number and form a series of fractions $P_n/Q_n$. For example, if

$$\rho = \cfrac{1}{a_1 + \cfrac{1}{a_2 + \cfrac{1}{a_3 + \cdots}}} = \langle a_1, a_2, a_3, \ldots \rangle \tag{5}$$

then the approximants to the golden mean $\rho_g = (\sqrt{5} - 1)/2 = \langle 1, 1, 1, \ldots \rangle$ are:

$$\langle 1 \rangle = \frac{1}{1}, \langle 1, 1 \rangle = \frac{1}{2}, \langle 1, 1, 1 \rangle = \frac{2}{3}, \ldots \tag{6}$$

The numbers $Q_n$ are necessary to define the scaling function, and for the golden mean rotation number they are the Fibonacci numbers ($Q_0 = 1$, $Q_1 = 2$, $Q_2 = 3$, $Q_n = Q_{n-1} + Q_{n-2}$). On the critical line, at the irrational winding number, one considers the first $Q_n$ points of the orbit. These points delimit segments $\Delta_s^{(n)}$ along the circle, grouped in a series of levels, indexed by $n$. For the distorted loop from the experiment we calculate the separation of points using arc length along the curve. For the sine circle map the angular separation is used. The ratio of these segments is used to define the scaling function. In figure 6 the first 13 points of the irrational golden mean orbit are used to delimit the segments at level 3. The segments $\Delta_s^{(n)}$ are found by iterating with the map the endpoints of segments $\Delta_0^{(n)}$. Because all points are obtained from the same orbit, the segments from previous levels can also be constructed from knowledge of the first 13 points, as indicated in figure 6. Notice that the ratio trick of reference [17] cannot be used, as all the different levels are computed at fixed parameters of the map, as required by the experiment.

The scaling function is built from a series of piecewise constant steps of height $\sigma_s^{(n)}$ placed in ascending order of the integer $s$ (which indexes the steps) and rescaled to span the unit interval. The scaling $\sigma_s^{(n)}$ is given by the ratio between the size of the segment $|\Delta_s^{(n+1)}|$ and its parent segment:

$$\sigma_s^{(n)} = \frac{|\Delta_s^{(n+1)}|}{|\Delta_{\Theta(s,Q_n)}^{(n)}|} \quad , \tag{7}$$

where $Q_n$ is the Fibonacci number of segments that are used at level $n$. The function $\Theta(s, Q)$ is the parent index function which in the simple case of the golden-mean returns $s - Q$ if $s \geq Q$ and $s$ otherwise. In figure 6, we have shown which segments to compare to compute the scalings $\sigma_s^{(0)}$, $\sigma_s^{(1)}$, and $\sigma_s^{(2)}$ at different levels. Formulas for other rotation numbers are given in the appendix of reference [7]. The scaling function is a function of the unit interval into itself. In terms of the step function $\theta$ ($\theta(x) = 1$ if $x > 0$ and 0 otherwise):

$$\sigma(t) = \sum_s \sigma_s^{(n)} \theta(t - t_s^{(n)}) \theta(t_{s+1}^{(n)} - t) \quad , \tag{8}$$

where the $t_s^{(n)}$ are where the discontinuities of the scaling function occur. The summation runs over a $Q_p$ Fibonacci number of them, and the value of $p$ depends on the level of approximation to the *universal* scaling function.

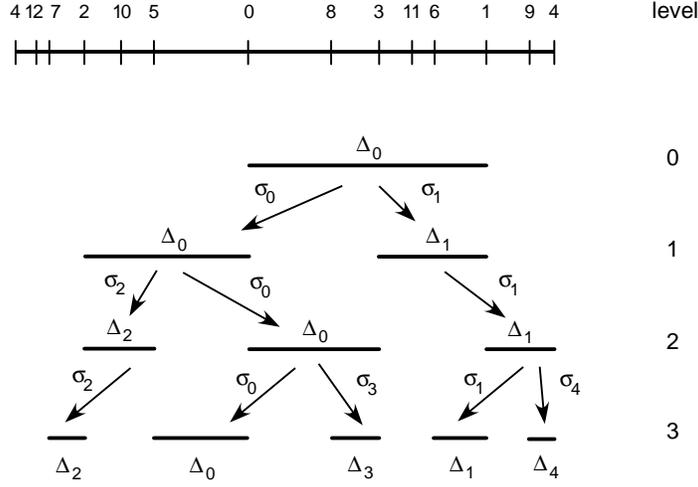

Figure 6: Arrangement of the segments used in constructing the scaling function. The endpoints of the segments $\Delta_k^{(n)}$ are the points indicated on the top line, and their ratios $\sigma_i^{(n)}$ are defined in equation 7. The levels $n$ are indicated to the side.

For the simple case of a five-step scaling function they are: $t_0^{(n)} = 0$, $t_1^{(n)} = Q_{n-3}/Q_n$, $t_2^{(n)} = Q_{n-2}/Q_n$, $t_3^{(n)} = Q_{n-1}/Q_n$, and $t_4^{(n)} = (Q_{n-3} + Q_{n-1})/Q_n$; the general case can be worked out by expanding $t$ in a Fibonacci basis. To obtain a universal scaling function two limits must be considered. First $n \to \infty$, in which the scalings $\sigma_s^{(n)}$ go to their limiting values. In this limit more and more of the points of the irrational orbit are considered, and the segments $\Delta_s^{(n)}$ get closer and closer to the inflection point of the circle map. Second $p \to \infty$, in which the number of terms of the sum goes to infinity. The first limit ($n \to \infty$) takes the scaling function to its universal form, whereas the second ($p \to \infty$) adds detail to the function. The limit towards detail cannot be taken before the limit towards universality. For experimental data the limit towards universality corresponds to considering a sequence of periodic orbits that approach the irrational one (tongue width going to zero), and the detailing limit corresponds to considering a larger number of levels.

One practical consequence of the double limit is that one cannot use the smallest possible region determined by a periodic orbit as the segments for the scaling function. From numerical simulations we observe that the convergence is improved if we use only the first $Q_j$ points of a $Q_k > Q_j$ periodic orbit in computing the ratios. For example, in an orbit of $Q_k = 17711$ if we compute the scaling function with segments at level 8 the first step of the five-step scaling function is at 0.48, close to its limiting value of 0.468, whereas if we compute the scaling function with segments at level 16 the same step is at 0.62, close to its trivial value. Notice that this implies that there are considerably fewer steps in the scaling function than would be expected from the experimental data set. In the scaling function we compute from experimental data we only consider the five-step scaling function. This is the largest number of scales we could extract given how closely we had approximated universality. As is illustrated in figure 7(a), the five-step theoretical scaling function is already a close approximation to the universal limiting function.

As criticality is approached the scaling function goes from trivial behavior to behavior that characterizes golden mean criticality. In any experiment the rotation number is not exactly a golden-mean tail and the amount of data is not infinite, so for an experimentally determined scaling function the transition from the trivial case to the critical case is smooth instead of abrupt. This is analogous to the transition of the $f(\alpha)$ spectrum [18]. The effects of finite orbits and inaccuracies in the control parameters can be seen in the scaling function. For a sub-critical orbit, if only the first few points of the orbit are used, then the scaling function resembles the critical sine circle map scaling function, but as more points of the orbit are considered the scaling function flattens out, creating the apparent contradiction that as more data is considered, less "accuracy" is obtained.

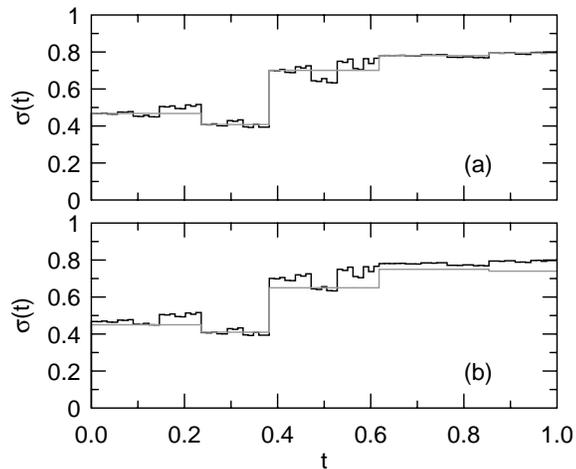

Figure 7: The numerical and experimental scaling functions compared to the limiting universal scaling function. In (a) the five-step numerical approximation in gray is laid over the limiting scaling function in black. The two functions agree at the left end of the large steps. In (b) the five-step experimental approximation in gray is laid over the same limiting function. The error bars for the experimental curve can be found in table 1.

In figure 8 we have plotted several five-scale scaling functions for the sine circle map for a sub-critical value of the control parameter. The scaling functions differ by the number of points considered from the orbit. If only a few initial points are considered the scaling function resembles the critical one, but as more of the orbit is taken into account, the non criticality of the map becomes evident. Similar behavior is observed with the experimental data. The scaling function obtained resembles the theoretical curve as long as the orbits are short, but as longer orbits are considered the experimental scaling function differs from the predicted one.

A similar phenomenon happens if we deviate from the superstable point along the critical line. In this case the scaling function is distorted away from the theoretical result, but still appears to be critical. In figure 9 we have plotted three scaling functions. One is for a rotation number smaller (below) the superstable point and the other is for a rotation number larger (above) than the superstable point. Notice that there can be large variations in the values of the scales. This points out the importance of being at exactly the superstable point in computing the scaling function.

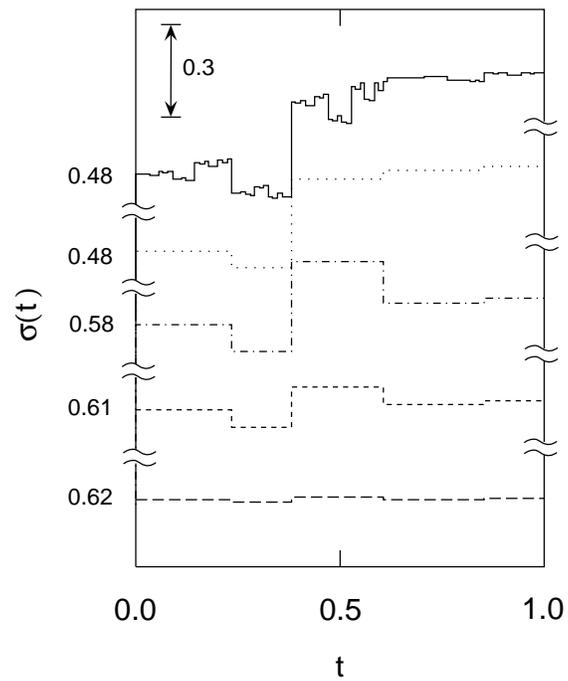

Figure 8: *Several scaling functions computed at a small distance from the irrational winding number. The topmost curve is the theoretical curve and the next uses only the first few points of the orbit, whereas the bottommost uses around a quarter of the points.*

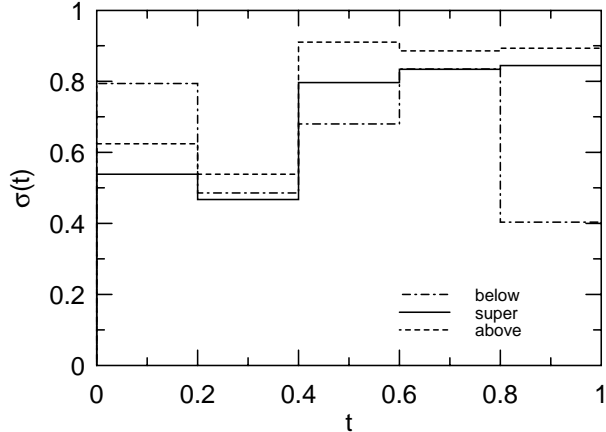

Figure 9: *Scaling functions along the away from the superstable point along the critical line. The plots show that large errors may occur when the system is not kept at the superstable point.*

A straightforward application of the scaling function definition would consist of choosing a large-$Q$ periodic orbit (a high order rational approximant to the golden mean) and using the smallest intervals at the largest level of an orbit to compute the scaling function. This disregards the order of the limits mentioned earlier. Such an approach was used previously, in combination with averaging, to reduce noise [2], and it fails because the convergence of ratios of intervals to their universal limit is nonuniform over the entire orbit and because averaging of noise does not improve uncertainty in the exact experimental winding number (in particular for drift in the control parameter, which we discuss later on). Explicit demonstration of the influence of noise on calculating the scaling function will be presented elsewhere [19].

Another problem with a straightforward application of the scaling function definition is that segments approach their universal limit differently for different sections of the periodic orbit. Because the universality of the quasiperiodic scaling function stems from the self-similarity of any map un-der composition and rescaling in the neighborhood of a cubic inflection point (the Cvitanović-Feigenbaum functional equation), convergence is most rapid in the vicinity of this point. This translates in the computation of the scaling function into considering only segments around the inflection point, that is, not all the segments around the circle can be used as indicated in figure 6. We accomplish this by considering the iterates of an initial segment not under the actual map $F$, but of one of its iterates, $F^{(k)} = F \circ \cdots \circ F$, composed $Q_k$ times. In figure 10 we have enlarged the region between the orbit points 0 and 8 of figure 6 and indicated the endpoints of the segments that would be used in computing the scaling function. The points of a 144/233 period orbit are plotted, but at most the first 89 points are used, which is three levels up from the bottom. This corrects the definition explained with the aid of figure 6.

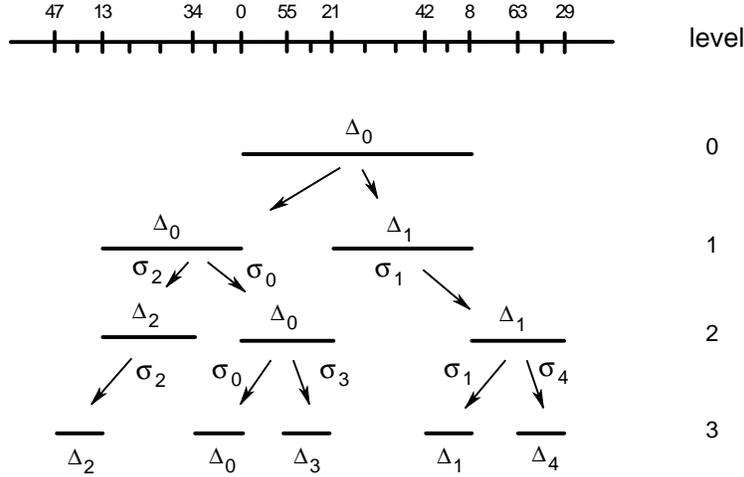

*Figure 10: Segments around the inflection point. This is an expanded view of the points around the segment from 0 to 8. The extra points indicated in the top line are from the first 144 points of the orbit (the 89/144 approximant). Notice that the arrangement of the segments is the same as in figure 6.*

In an experiment the circle map is not given analytically, and there is no direct way of determining the inflection point. We use the empirical criterion of Wang *et al.* [10]. It consists of observing that the inflection point, if it exists, is an endpoint of $\Delta_0^{(n)}$ which should be the largest segment at a given level. Although it is not essential that the segment containing the inflection point be used as $\Delta_0^{(n)}$, the convergence is fastest if the segments are organized around it. Plotting the return map or its derivative is not a method for determining the inflection point, as it may be absent in a map within an Arnold tongue [10, 20].

The final point that must be considered is experimental in nature and does not arise in numerical simulations. In order to obtain a reasonable scaling function one must be at a superstable orbit within a mode-locked tongue (in practice close to the middle of the tongue). In an experiment, the data set closest to the irrational winding number at the critical line is usually chosen. The closer the system is to the irrational winding number, the longer the orbit that can be obtained and the longer its control parameters must be kept locked within an Arnold tongue. But the width of a tongue decreases as the period of the orbit increases, and the longest orbit is obtained when the tongue width is below the resolution of the apparatus. In practice, for the narrowest tongues the system will jump between several mode-locked states as the control parameters are kept fixed within experimental resolution. To avoid this instability in the mode-locked state, orbits of shorter period should be considered. The experimentalist is then faced with the choice of either stable, short, and less converged orbits; or fluctuating, long, and better converged orbits. The reconstruction of the dynamics [21], described next, achieves an optimal compromise between these constraints and allows significant improvements over the direct evaluation of the scaling ratios.

To reconstruct the dynamics we determine two data sets that are close by in parameter space and on opposite sides of the golden-mean rotation number. We proceed to determine an interpolated map from each finite set of experimental points using a least-squares cubic spline fit which is then iterated to determine its winding number. We then interpolate between the two maps to obtain the superstable point within one of the intermediate mode-locked tongue. (Because the two maps used for interpolation are close together, we use linear interpolation between their ordinates.) As a consistency check on our interpolation scheme we have computed the rate of contraction $\delta$ of the Arnold tongues as the golden mean is better approximated. It is measured to be 2.8, to be compared with the prediction of 2.83 from the renormalization group for the circle map [4].

From the superstable interpolated map the five-step scaling function is computed. With the interpolation method we can determine periods of lengths limited only by the computer, but we have been careful not to use periods that lead to average segment sizes that are smaller than the segments from the data. If this precaution is not taken the method will generate orbits whose universality class is dictated by the nature of the interpolating spline. The scaling function of the map reconstructed from the data is given in table 1 and plotted in figure 7(b). For comparison the theoretical scaling ratios for the universality class of the sine circle map, and the scaling function computed without the reconstruction process are also given. The theoretical scaling ratios were computed from a 832040/1346269 orbit of the sine circle map. The tabulated reconstructed scaling function is not the result of averaging over several data sets, but computed from a single long orbit. The errors are estimated based on several different rotation numbers with golden mean behavior. The effects of not being exactly at the irrational rotation number can be seen in the direct scaling function: the first three scales are in good agreement, but the final two, where small errors have accumulated, are not.

| scale | limiting | experimental | direct |
|---|---|---|---|
| $\sigma_0$ | 0.468 | $0.45 \pm 0.04$ | 0.42 |
| $\sigma_1$ | 0.407 | $0.41 \pm 0.05$ | 0.36 |
| $\sigma_2$ | 0.700 | $0.65 \pm 0.03$ | 0.70 |
| $\sigma_3$ | 0.781 | $0.75 \pm 0.03$ | 0.54 |
| $\sigma_4$ | 0.794 | $0.74 \pm 0.05$ | 0.59 |

Table 1: Five step scaling functions: limiting value, experimental value obtained with reconstruction, and directly without reconstruction.

## 5 Conclusions

What have we gained in our analysis relative to, for example, computing an $f(\alpha)$ spectrum? First we know that the dynamics are correct because the construction of the scaling function requires constructing the symbolic dynamics of the map whereas the spectrum does not distinguish between fractal sets with the same statistics but different dynamics [22]. Second, we have extracted three scales beyond the $f(\alpha)$ spectrum [23]. Thus by extracting five scaling ratios that agree within 10% with the theoretical predictions we have made the most stringent test to date of quasiperiodic universality. In summary, it is possible to extract a scaling function from experimental data only if the orbit points around the inflection point are considered and if the parameters of the system are adjusted to be at the irrational winding number. Experimentally both constraints are interconnected and can be resolved by reconstructing the dynamics.

We acknowledge useful conversations with P. Cvitanović, M. Feigenbaum, J. Lowenstein, T. Sullivan, and C. Tresser. This work was funded by the U.S. Department of Energy, Basic Energy Science, Division of Materials Science.